\documentclass[amsmath,amsfonts,amssymb,twocolumn,superscriptaddress,showpacs]{revtex4}

\usepackage{graphicx,psfrag,color}
\usepackage{dcolumn}
\usepackage{bm}
\usepackage{mathbbol}

\def\Z{{\mathbb{Z}}}
\def\one{{\mathbb{1}}}

\begin{document}

\title{Bond Algebras and Exact Solvability of Hamiltonians: \\
Spin $S$=1/2 Multilayer Systems and Other Curiosities}

\author{Zohar Nussinov}
\affiliation{Department of Physics, Washington University, St.
Louis, MO 63160, USA}
\author{Gerardo Ortiz}
\affiliation{Department of Physics, Indiana University, Bloomington,
IN 47405, USA}

\date{\today}

\begin{abstract}
We introduce an algebraic methodology for designing exactly-solvable Lie
model Hamiltonians. The idea consists in looking at the algebra
generated by {\it bond operators}.  We illustrate how this method can be
applied to solve numerous problems of current interest in the context of
topological quantum order. These include Kitaev's toric code and
honeycomb models, a vector exchange model, and a Clifford $\gamma$ model
on a triangular lattice.
\end{abstract}

\pacs{05.30.-d, 03.67.Pp, 05.30.Pr, 11.15.-q}
\maketitle


\section{Introduction}

Whenever one is interested in studying a new physical phenomena whose 
effective model includes degrees of freedom (spins, fermions, bosons,
etc.) which are strongly coupled, one attempts to invoke simplifying
assumptions hoping that the resulting problem represents the relevant
minimal model for the phenomenon at hand. Finding exactly-solvable
models is always welcome.  This paper is about a general methodology to
generate exactly-solvable Hamiltonians  by concentrating on the algebra
generated by algebraic objects called {\it bonds}. We have already used
this methodology in Ref. \cite{orbit}, where we solved a doped orbital
compass model in two and three space dimensions, although we did not
explained the generality of the mathematical approach. A goal of this
paper is to present this  methodology in full  detail  and  show that by
using the algebra of bond operators one can  easily construct whole
families of exactly-solvable models, several of these displaying
topological quantum order \cite{wenbook}, and spin liquid behavior.  

For the sake of clarity, we  will focus on quantum lattice systems which
have $N_s= \prod_{\mu=1}^{D} L_{\mu}$  sites, with $L_\mu$ the number of
sites along each spatial direction $\mu$, and $D$ the  dimensionality of
the lattice. The connectivity of the lattice and its general {\it
topology} are of paramount importance.  Associated with each lattice
site ${\bf i} \in \Z^{N_s}$ there is a Hilbert space ${\cal H}_{\bf i}$
of finite dimension ${\cal D}_{\bf i}$. The total Hilbert space is the
tensor product of the local state spaces, ${\cal H} = \bigotimes_{\bf i}
{\cal H}_{\bf i}$, in the case of distinguishable subsystems (or a
proper subspace in the case of indistinguishable ones), and its
dimension is ${\cal D}=\prod_{{\bf i}=1}^{N_s}{\cal D}_{\bf i}$.

Let us first start with an intuitive introduction to the key concept of
{\it bond algebras}. We consider situations in which the Hamiltonian of
a system $H$, whose state space is ${\cal H}$, can be written as a sum of
{\it quasi-local} terms or {\it bonds} ($\{h_{R}\}$),
\begin{eqnarray}
H = \sum_{R} \alpha_R h_{R} ,
\label{Hg}
\end{eqnarray}
where $\alpha_R$ is a $c$-number ($H$ must be an Hermitian operator) and
$R$ includes a finite number of lattice sites $\bf i$.  In general, the
operators $h_{R}$ will generate a certain ({\it bond}) algebra $\cal G$
whose dimension is $O({\cal D})$. To simplify the description, in the
following, we are going to concentrate on semisimple Lie algebras.
Notice that we do not constrain ourselves to a particular representation
of the algebra. 

It may happen that the Hamiltonian itself is an element of a subalgebra
of $\cal G$ of dimension poly$\log {\cal D}$. If such is the case, $H$, 
which represents a {\it Generalized Mean-Field Hamiltonian} (GMFH),  is
{\it exactly-solvable} \cite{solvability}, and there is a polynomially
in $\log {\cal D}$ efficient algorithm to diagonalize it
\cite{solvability}. We say that $H$ is exactly-solvable when an
arbitrarily chosen eigenvalue, and an appropriate description of the
corresponding eigenstate, can be obtained and represented to precision
$\epsilon$ by means of a classical algorithm efficient in $\log {\cal
D}$ and $1/\epsilon$. This definition, motivated by complexity theory,
yields a sufficient criterion for exact-solvability. A particular case
of exact solvability is when the spectrum can be expressed in closed
form. 

A main contribution of this paper is to propose a methodology to
generate such Hamiltonians by using two mathematical principles that
will become evident in the following sections. In all cases, this
methodology rests on  (1) Topological constraints that are related to
the connectivity of the lattice  Hamiltonian or graph. In several
instances, it further relies on the existence of (2) gauge symmetries.
These symmetries allow a decomposition of the  Hilbert space into
sectors. The operators $\{h_R\}$  belong to the lowest dimensional
representation of the algebra on the Hilbert space, or its sub-spaces.
In the following we illustrate the bond algebra methodology by showing 
some tutorial examples of known trivially {\it exactly-solvable}
problems. 

\subsection{Ising model}
\label{Isingtut}

A simple example is afforded by the Ising model on a hypercubic lattice
of $N_s$ sites, 
\begin{eqnarray}
H_{\sf Ising} = - \sum_{\langle i j \rangle } J \sigma_{i} \sigma_{j}.
\label{Ising}
\end{eqnarray}
The bonds $b_{ij} \equiv \sigma_{i} \sigma_{j}$ satisfy a simple Ising
(Abelian) type algebra defined on a ${\cal D}=2^{N_s}$-dimensional space
(the span of the original Ising system):
\begin{eqnarray}
[b_{ij}, b_{kl}]=0, ~~~~b_{ij}^{2} =1 , 
\end{eqnarray}
since $\sigma_i=\pm 1$.  All classical Hamiltonians are extreme cases of
GMFHs: Its spectra are trivially determined. 

\subsection{Transverse field Ising chain}
\label{strfm}

The Hamiltonian of a single transverse field Ising chain of length $N_s$
reads
\begin{eqnarray}
H_{\sf TFIM} = - \sum_{i=1}^{N_s}  ( J_i \sigma^{y}_{i} \sigma^{y}_{i+1}
+ h_i \sigma^{x}_{i} ),
\label{TFIM}
\end{eqnarray} 
where $\sigma_{i}^{\mu}$ ($\mu=x,y,z$) represent Pauli matrices. To make
clear the algebraic connection that will follow, let us denote the two
terms (transverse field and bond variables) as follows:
\begin{eqnarray}
\bar{A}_{i}^{x} = \sigma^{x}_{i}, \bar{A}_{i,j} = \sigma^{y}_{i}
\sigma^{y}_{j}.
\label{bdt}
\end{eqnarray}
In terms of these, the Hamiltonian of Eq.(\ref{TFIM}) obviously reads
\begin{eqnarray}
H_{\sf TFIM} = - \sum_{i=1}^{N_s}  ( J_i \bar{A}_{i,i+1}+ h_i
\bar{A}_{i}^{x} ),
\label{AAt}
\end{eqnarray}
with interaction terms satisfying
\begin{eqnarray}
[\bar{A}^{x}_{i}, \bar{A}^{x}_{j}]&=&0= 
[\bar{A}_{i,j},  \bar{A}_{k,l}] \nonumber \\ 
\{\bar{A}^{x}_{i}, \bar{A}_{i, i+1}\}&=&0=\{\bar{A}^{x}_{i},
\bar{A}_{i-1,i}\},\nonumber  \\
{[}\bar{A}^{x}_{i}, \bar{A}_{j,k}] &=&0 \ , \ i \neq j,k \nonumber  \\
(\bar{A}^{x}_{i})^{2} &=&1=(\bar{A}_{j,k})^{2} ,
\label{condt1}
\end{eqnarray}
which forms an so$(2N_s)$ (poly$\log {\cal D}$) algebra with ${\cal
D}=2^{N_s}$.

Note that the bond algebra encapsulated in the relations above is
invariant under the flip of any transverse field locally. The
transformation
\begin{eqnarray}
\bar{A}_{i}^{x} \to -\bar{A}_{i}^{x}
\label{hflip}
\end{eqnarray}
effects $h_{i} \to - h_{i}$ at the lattice site $i$. Indeed, all that a
flip of local fields does is to leave the spectrum unaltered while
permuting the eigenstates amongst themselves.  In more conventional
terms, the invariance of the spectrum mandated by the invariance of the
bond algebra under the transformation of Eq.(\ref{hflip}) is seen by
noting that a similarity transformation with the local unitary (and
Hermitian) operator $U_{i} = \sigma^{y}_{i}$ sets
\begin{eqnarray}
\sigma^{y}_{i} \sigma^{x}_{i} \sigma^{y}_{i} = -  \sigma^{x}_{i}
\end{eqnarray}
while leaving $\sigma^{y}_{i}$ and thus $\bar{A}_{i,i+1}$ invariant. The
spectrum of Eq.(\ref{AAt}) can be determined by performing a
Jordan-Wigner transformation to free fermions. Equivalently,  it may
noted that the bond algebra of a tight-binding spinless Fermi model
(with pairing terms) is equivalent to that of Eq.(\ref{condt1}). 

\subsection{Orbital compass chain model}

This model was introduced in \cite{oles}. It consists of a $D=1$
dimensional system with alternating  $xx$ and $yy$ interactions. Namely,
consider a chain of length $N_s$ in which the Hamiltonian is given by 
\begin{eqnarray}
H_{\sf OCM} = \sum_{i=1,3,5,\cdots} J_{x,i} \sigma_{i}^{x}
\sigma_{i+1}^{x}  + \sum_{i=2,4,6,\cdots} J_{y,i} \sigma_{i}^{y}
\sigma_{i+1}^{y}.
\label{1docm}
\end{eqnarray}

Let us define the even and odd bonds by
\begin{eqnarray}
A_{m} = \sigma_{2m}^{y} \sigma_{2m+1}^{y} \ , \ 
B_{m} = \sigma_{2m-1}^{x} \sigma_{2m}^{x}.
\end{eqnarray}
They satisfy the following algebra (${\cal D}=2^{N_s}$)
\begin{eqnarray}
[A_m, A_n]&=&0= [B_m,B_n] \nonumber \\ 
\{A_{m}, B_{m}\}  &=&0= \{A_{m}, B_{m+1}\},\nonumber \\
{[}A_{m}, B_{n}] &=&0 \ , \  |m-n|>1 \nonumber \\
(A_m)^{2} &=&1=(B_m)^{2} .
\end{eqnarray}
This algebra is identical to the algebra of bonds of Eqs.(\ref{condt1}).
For an open chain, there are no boundary conditions on the bonds in
either problem. If we enabled interactions (both exchange and transverse
fields) on only one half of the chain (that is, if the sum in
Eq.(\ref{TFIM}) would extend, for even $N_s$, only from $ 1 \le i \le
N_s/2$) and add $N_s/2$ non-interacting spins, then the number of
interaction terms in Eq.(\ref{TFIM}) and Eq.(\ref{1docm}), their
algebras (and dimension of their representations), and the size of the
Hilbert space on which both systems are defined are identical. In that
case, the partition functions are identical up to a trivial
multiplicative factor (after identifying $J_i=J_{y,i}$ and
$h_i=J_{x,i}$)
\begin{eqnarray}
{\cal Z}_{\sf OCM}(N_s)  = 2^{N_s/2} {\cal Z}_{\sf TFIM}(N_s/2).
\end{eqnarray}
Such a relation was indeed found by \cite{oles}  by an explicit
diagonalization of the Fermi bilinear found after a Jordan-Wigner
transformation performed on $H_{\sf OCM}$. Here we arrived at the same
result by a trivial application of the methodology of bond algebras. 

\subsection{Kitaev's Toric Code Model}
\label{k2}

Kitaev's toric code model \cite{kitaev} is defined on a square lattice
with $L \times L=N_s$ sites, where on each bond (or link) $(ij)$ is an
$S=1/2$ degree of freedom indicated by a Pauli matrix
$\sigma_{ij}^{\mu}$. The Hamiltonian acting on a ${\cal D}=
2^{2N_s}$-dimensional Hilbert space is
\begin{eqnarray}
H_{K} = -\sum_{s} A_{s} -\sum_{p} B_{p}
\label{kitaevmodel}
\end{eqnarray}
with Hermitian operators (whose eigenvalues are $\pm 1$)
\begin{eqnarray}
A_{s} = \prod_{(ij) \in {\sf star}(s)} \sigma_{ij}^{x}, 
~~B_{p} = \prod_{(ij) \in {\sf plaquette}(p)} \sigma_{ij}^{z} ,
\label{AB_defn}
\end{eqnarray}
where $B_{p}$ and $A_{s}$ describe the plaquette (or face) and star (or
vertex) operators associated with each plaquette $p$, and each site $s$
of the square lattice. The reader may want to consult Refs.
\cite{NO1,NO2,no_p} for notation purposes.

That the $D=2$ Kitaev's toric code model is identical to two decoupled
Ising chains \cite{NO1,NO2} is immediately seen by looking at the bond
algebra. The algebra of the bonds given by Eq.(\ref{AB_defn}) is
trivial, it is an  Ising (Abelian) type algebra
\begin{eqnarray}
[A_{s}, B_{p}]&=& [A_{s}, A_{s'}]= [B_{p}, B_{p'}]=0, \nonumber \\ 
(A_{s})^{2} &=&1= (B_{p})^{2} . 
\label{algkit}
\end{eqnarray}
For periodic boundary conditions one has the additional constraint
\begin{eqnarray}
\prod_{s} A_{s} =\prod_{p} B_{p} = 1.
\label{conkit}
\end{eqnarray}

It is very easy to realize that the Hamiltonian for two decoupled
Ising chains, each of length $N_s$
\begin{eqnarray}
H_{I} = - \sum_{s=1}^{N_s} \sigma_{s} \sigma_{s+1}  - \sum_{p=1}^{N_s}
\tau_{p} \tau_{p+1}
\label{h1d}
\end{eqnarray}
with $\sigma_{s} = \pm 1$ and $\tau_{p} = \pm 1$, displays an identical
bond algebra to Eqs.(\ref{algkit}), with the same representation.  
Thus, one can immediately write down the partition function
\cite{NO1,NO2,no_p}
\begin{eqnarray}
{\cal Z}_K = (2\cosh\beta)^{2N_s} (1+ \tanh^{N_s} \beta)^{2},
\label{zend}
\end{eqnarray}
where $\beta=1/(k_B T)$, and $T$ is temperature. Moreover, the bond
algebra of Kitaev's toric code model is identical to that of Wen's
plaquette model \cite{wenmodel} which proves the equivalence of the two
systems \cite{NO2}.  It is worthwhile to note that Eq.(\ref{zend}) is
also the outcome of a high-temperature series expansion  \cite{longT}.

Thus, Kitaev's toric code model is identical to a one-dimensional Ising
system. This statement has ramifications for the stability of  quantum
memories --- an item that we investigated in detail early on
\cite{NO1,NO2,no_p}. This mapping allows not only an evaluation of the
partition function but also a direct computation of all correlators in
Kitaev's toric code model. For a detailed explanation see Refs.
\cite{NO2, no_p}. In particular, see  subsections (XIII A,B) as well as
footnotes [61-63] of Ref. \cite{NO2}. The equations of motion with
uncorrelated noise are insensitive to a change of basis. Consequently,
the dynamics and thermal effects present in one-dimensional systems rear
their head also in Kitaev's toric code model. In particular, the system
is unstable to thermal noise --- a phenomenon that we coined {\it
thermal fragility} \cite{NO1, NO2, no_p} and has been recently confirmed
by others \cite{kc, afh, afh7, st}.  Our bond algebraic mapping enables
an immediate extraction of crossovers for finite size systems
\cite{crossover_trivial}. 

\subsection{Plaquette model in a transverse magnetic field}

This model \cite{wenfield} is defined on a square lattice as follows:
\begin{eqnarray}
H_\Box= - \sum_i J_i F_{i}  - \sum_i h_i \sigma_{i}^{x} ,
\label{went}
\end{eqnarray}
where $F_{i} = \sigma^{x}_{i} \sigma^{y}_{i+ \hat{e}_{x}}
\sigma^{x}_{i+\hat{e}_{x}+\hat{e}_{y}} \sigma^{y}_{i+\hat{e}_{y}}$, with
$\hat{e}_{\mu}$ representing unit vectors along the $\mu$ direction in
the lattice. 

Setting, $G_{i} \equiv \sigma^{x}_{i}$, and $F_{i} \equiv F_{i*}$ with
$i* \equiv i+ \frac{1}{2}(\hat{e}_{x} + \hat{e}_{y})$, and a diagonal
chain coordinate $j $ along the (1,1) direction, that alternates between
$i$ and $i*$, the algebra of the interaction terms ({\it bonds})  in
Eq.(\ref{went}) is
\begin{eqnarray}
{[}F_{j}, F_{j'}{]} &=&0= [G_{j}, G_{j'}] \nonumber \\ 
\{F_{j},G_{j+1}\} &=&0=\{F_{j},G_{j-1}\},\nonumber  \\
{[}F_{j}, G_{j'}{]}&=&0~~~ (j-j' \neq \pm \frac{1}{2} (\hat{e}_{x} +
\hat{e}_{y})) \nonumber  \\
(F_{i})^{2} &=&1=(G_i)^{2} .
\end{eqnarray}
For a system with open boundary conditions, the algebra of this system
is none other than that of a stack of decoupled  transverse field Ising
chains (see subsection \ref{strfm} and  Eq.(\ref{condt1}) in
particular)  all oriented diagonally along the (1,1) direction. 
Setting, in Eq.(\ref{condt1}), $\bar{A}_{ij} \equiv \bar{A}_{i*}$, we
see that the algebra and the dimension of the Hilbert space in both
problems are identical. Indeed, a more elaborate treatment finds that
this system is none other than that of a transverse field Ising model
\cite{wenfield} precisely  as we find by examining the bond algebra. The
partition function is, therefore, exactly the same as that of a
transverse field Ising model.

In the next sections, we illustrate the power of our method by reviewing
several, more challenging, known examples of exactly-solvable models
whose solutions can be immediately achieved in this way,  and then we
turn to new models that we introduced and solved using these tools. We
start by discussing Kitaev's honeycomb model \cite{Kitaev2006} and show
that no enlargement of the Hilbert space \cite{Kitaev2006} nor a direct
Jordan-Wigner mapping \cite{nussinov_chen} is neccessary to solve this
model in a very short and direct manner.  Next, we will turn to new
models. The first of these is the {\it vector exchange} model which
forms a simple extension of Kitaev's honeycomb model. We will later on
show that all these models have in common a Clifford algebraic structure
on-site and an Abelian structure off-site. This defines a simple class
of GMFH of the so$(2N)$ type \cite{solvability}. More general Lie
algebraic structures can be also realized. 

\section{Kitaev's Honeycomb model}

\subsection{Spectrum from bond algebras}
\label{alki}

Kitaev's honeycomb lattice model \cite{Kitaev2006, bss} is a member of a
family of models whose Hamiltonians  are elements of the so$(2N_s)$
algebra, where $N_s$ is the number of vertices of the honeycomb lattice,
i.e., it is a GMFH.

\begin{figure}[h]
\includegraphics[width=3.4in]{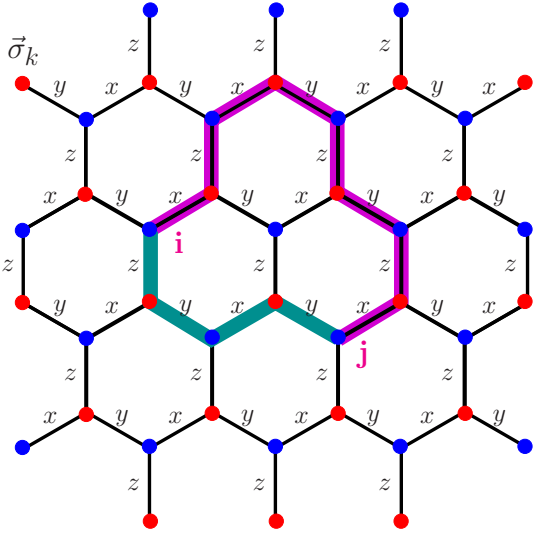}
\caption{Kitaev's model on a honeycomb lattice and three types of
bonds. On each vertex there is an $S=1/2$ degree of freedom indicated
by a Pauli matrix $\vec{\sigma}_{k}$. There are two different types of 
vertices.  Thick-colored contours represent arbitrary
paths drawn on the lattice, e.g., from site $i$ to $j$.}
\label{fig1}
\end{figure}

The model is defined by the following $S=1/2$ Hamiltonian (Fig.
\ref{fig1})
\begin{eqnarray}
H_{K_h}\!\!&=&\!\!-J_x \!\!\!\!\sum_{x{\sf-bonds}}
\!\!\!\sigma^x_{i}\sigma^x_{j} -J_y\!\!\!\!\sum_{y{\sf -bonds}}
\!\!\!\sigma^y_{i}\sigma^y_{j} -J_z\!\!\!\!\sum_{z{\sf -bonds}}\!\!\!
\sigma^z_{i}\sigma^z_{j} \nonumber \\ 
&=& - \sum_{\langle ij \rangle} J_{ij} \sigma_{i}^{\mu} \sigma_{j}^{\mu}
~~~  (\hat{e}_{\mu} || (\vec{j} - \vec{i})).
\label{H}
\end{eqnarray}

Let us define the bond operators
\begin{eqnarray}
A_{ij}^\mu=\sigma^\mu_i \sigma^\mu_j  \ , \ \mu=x,y,z ,
\label{hann}
\end{eqnarray}
where $\langle i,j\rangle$ defines a bond (nearest-neighbors). 
We can assign a uniform direction to all bonds 
by choosing for all nearest neighbor sites $i$ and $j$,
the bond defined by
$A_{ij}^{\mu}$ corresponds to $j-i = \hat{e}_{\mu}$
(and not $j-i = - \hat{e}_{\mu}$).
From the commutation relations for  $SU(2)$ spins it is clear that
($\Delta_{ijkl}=\delta_{ik}+ \delta_{jk}+\delta_{il}+\delta_{jl}$)
\begin{eqnarray}
[A_{ij}^\mu,A_{kl}^\nu]=(1-\delta_{\mu\nu})(1-(-1)^{\Delta_{ijkl}}) \
A_{ij}^\mu A_{kl}^\nu ,
\end{eqnarray}
and, moreover, it is clear that $(A_{ij}^\mu)^2=1$. For an arbitrary set
of bond operators $A_{ij}^\mu$, the Lie algebra $\cal G$ generated is
$O({\cal D})$. However, the set of bond operators that appear in 
$H_{K_h}$, $\{A_{ij}^\mu\}_{H_{K_h}}$, forms a Lie subalgebra of
$\cal G$, because of the particular lattice topology: Bond operators
that share a vertex anticommute, otherwise they commute. This subalgebra
is precisely so$(2N_s)$, and the Hamiltonian being an element of that
subalgebra is an GMFH. 

One could, in principle, stop here and diagonalize the problem in a
Hilbert space of dimension ${\cal D}=2^{N_s}$. However, there is a
further simplification in this problem. The simplification is related to
exploting the existence of {\it gauge symmetries}. Consider the {\it
anyon charge} \cite{Kitaev2006} operators
\begin{eqnarray}
I_{h_{\alpha}} = \prod_{\langle i j \rangle \in h_{\alpha}}
A_{ij}^\mu \nonumber = \sigma^{x}_{1}
\sigma^{y}_{2} \sigma^{z}_{3}  \sigma^{x}_{4} \sigma^{y}_{5}
\sigma^{z}_{6}
\label{iha}
\end{eqnarray}
where $h_\alpha$ defines a particular hexagonal plaquette (see Fig.
\ref{fig1}). The $\{I_{h_{\alpha}}\}$ operators have eigenvalues 
$\tilde{I}_{h_{\alpha}} = \pm 1$, and they satisfy the following
relations
\begin{eqnarray}
{[}I_{h_{\alpha}}, I_{h_{\alpha'}}] &=& 0 \ , \  (I_{h_{\alpha}})^2=1 ,
\nonumber \\
{[}I_{h_{\alpha}}, A_{ij}^\mu]&=& 0,
\end{eqnarray} 
which implies that $[I_{h_{\alpha}},H_{K_h}]=0$. In other words, the
product of bonds  taken around the hexagon $h_\alpha$ in a uniform
orientation (either clockwise or counter-clockwise) is a gauge symmetry.
The $N_{s}/2$ operators $\{\tilde{I}_{h_{\alpha}}\}$ satisfy the global
constraint
\begin{eqnarray}
\prod_{h_{\alpha}} \tilde{I}_{h_{\alpha}}=1.
\end{eqnarray}

This allows us to decompose the $\cal D$-dimensional Hilbert space 
$\cal H$ into $2^{N_s/2-1}$ orthogonal Hilbert subspaces ${\cal
H}_\eta$, each of dimension $\dim({\cal H}_\eta) = 2^{N_s/2+1}$ 
\begin{eqnarray}
{\cal{H}} = \bigoplus_{\eta=1}^{2^{N_s/2-1}} {\cal{H}}_{\eta}
\label{Hilbert}
\end{eqnarray}
Each Hilbert subspace ${\cal{H}}_{\eta}$ is characterized by a
particular set of eigenvalues $\{\tilde{I}_{h_{\alpha}}\}$ and 
projector
\begin{eqnarray}
\hat{P}_\eta =  \prod_{\alpha=1}^{N_s/2} \frac{\one +
\tilde{I}_{h_{\alpha}} I_{h_{\alpha}}}{2} = \prod_{\alpha=1}^{N_s/2}
\hat{P}_{I_{h_{\alpha}}}.
\label{proj}
\end{eqnarray}
The algebra satisfied by the projected bond operators
$\bar{A}_{ij}^\mu=\hat{P}_\eta{A}_{ij}^\mu\hat{P}_\eta$ is so$(2N_s)$,
but acts on a Hilbert (carrier) subspace of dimension $2^{N_s/2+1}$

To determine the spectrum in each subspace we look for an {\it
oscillator realization} of the algebra
\begin{eqnarray}
\bar{A}_{ij}^\mu= 2i \eta_{ij}c_i c_j \ , \ \mu=x,y,z ,
\label{ha}
\end{eqnarray}
in terms of Majorana fermions $c_i$, which satisfy
\begin{eqnarray}
\{c_i,c_j\}=\delta_{ij} \ , \ c_i^\dagger =c_i^{\;} .
\end{eqnarray}

The smallest representation of $N_s$ Majorana fermion modes ($N_s$ even)
is that in a $2^{N_s/2}$ dimensional Hilbert space. For reasons that
will become clear later on,  in Eq.(\ref{ha}) we will set $\eta_{ij}
=1$  on all bonds parallel to the ``$x$'' or ``$y$'' directions and
allow  $\eta_{ij} = \pm 1$ on all vertical bonds (those parallel to the
``$z$'' direction in which $\vec{j} - \vec{i} = \pm \hat{e}_{z}$). 
There remains an additional degeneracy factor of two (the  Hilbert space
dimension is of size $2^{N_s/2+1}$  while the representation of the
bonds is on a Hilbert space of size $2^{N_s/2}$). 

It is straightforward to show that the bilinear combinations of Majorana
fermions satisfy the same algebra and also constraints as the algebra of
$\{A^\mu_{ij}\}_{H_{K_h}}$. Notice that any connected open string product
of bonds becomes a bilinear in Majorana fermions \cite{nussinov_chen}
(see Fig. \ref{fig1})
\begin{eqnarray}
S_{i_1,i_L}\!\!=\bar{A}_{i_1i_2}^{\mu_1}\bar{A}_{i_2i_3}^{\mu_2} \cdots
\bar{A}_{i_Li_{L+1}}^{\mu_L}\!\!= 2 i^L \Big( \prod_{\langle i,j \rangle}
\eta_{ij}  \Big) \ c_{i_1} c_{i_{L+1}} . 
\label{stringt}
\end{eqnarray}
It turns out that all open strings having the same end points $i_1,i_L$
and with alternating $\mu's$ (e.g. $x,z,y,z,y,x,z,x$) can be of only
4 types 
\begin{eqnarray}
S_{i_1,i_L}=(\pm 1, \pm i) \ 2 c_{i_1} c_{i_{L+1}} ,
\end{eqnarray}
and form a polynomial in the number of vertices (or bonds) Lie algebra. 
The correspondence between the anyon charge sector
$\{\tilde{I}_{h_\alpha}\}$ and the set $\{\eta_{ij}\}$ is 
\begin{eqnarray}
\prod_{\langle ij \rangle \in h_{\alpha}} \eta_{ij}=
\tilde{I}_{h_{\alpha}}.
\label{gah}
\end{eqnarray}

The set $\{h_{\alpha}\}$ spans all fundamental hexagonal plaquettes from
which all closed loops $\Gamma$ can be uniquely constructed. The number
of plaquettes $\{h_{\alpha}\}$ is given by half the number of sites
$N_s/2$ as is the number of vertical bonds on which we may asign one of
the two phases corresponding to $\eta_{ij} = \pm 1$. With the
identification of Eq. (\ref{gah}), both sides of Eq. (\ref{ha}) satisfy
the same set of algebraic relations in each of the $2 \times 2^{N_s/2}$
dimensional Hilbert subspaces. 

This mapping allows us to immediately write down the spectrum in each
sector of $\{I_{h_\alpha}\}$ and to reproduce the results of
\cite{Kitaev2006,nussinov_chen} without the need for introducing two
Majorana fermions per spin and then  projecting out one (as in
\cite{Kitaev2006}) nor writing expliciting a Jordan-Wigner
transformation between fermionic and the spin variables (as was done in
\cite{nussinov_chen}). 

In a given sector $\eta=\{\eta_{ij}\}$, we have the Majorana fermion
representation of the Hamiltonian,
\begin{eqnarray}
H_{K_h,\eta} =2 i \sum_{\langle ij \rangle} \eta_{ij} J_{ij} c_{i} c_{j}
\label{he}
\end{eqnarray}
where $J_{ij} = J_{x}$ if $i$ and $j$ are separated by an ``$x$'' type
bond.  Similarly, $J_{ij} = J_{y,z}$ if  $i$ and $j$ are linked by a
``$y$'' or ``$z$'' type bond. Within the ground state sector 
($I_{h_\alpha} =1$ for all plaquettes $h_\alpha$), we may set
$\{\eta_{ij} =1\}$  and obtain the quasi-particle spectrum, $\vec{k}=(k_x,k_y)$
\cite{Kitaev2006,nussinov_chen}, 
 \begin{eqnarray} 
E_{\vec{k}} &=& \pm \sqrt{\epsilon_{\vec{k}}^2+\Delta_{\vec{k}}^2}, \nonumber \\ 
\epsilon_{\vec{k}}&=&2J_z-2J_x\cos k_x-2J_y\cos k_y, \nonumber \\ 
\Delta_{\vec{k}} &=&2J_x\sin k_x+2J_y\sin k_y.
\label{eD}
\end{eqnarray}

Our mapping allows for a closed-form solution only for a reduced set of
sectors (such as the ground state sector). For the rest, we still can
compute each eigenvalue and eigenvector with polynomial in $N_s$
complexity by using the Jacobi method \cite{solvability}. Thus, it is
not simple to compute the partition function of the model with the same
complexity. One can write down a formal solution, as we discuss below, 
but it is not a closed-form analytical solution in terms of simple
functions. 

\subsection{Partition function}

Although many results appear on the zero temperature behavior of
Kitaev's honeycomb model, there are very few results at finite
temperatures.  An exception is Ref. \cite{az}  which provides a finite
temperature metric analysis of Kitaev's honeycomb model. Related
results are discussed in \cite{no_p}.

The partition function includes contributions from all sectors and reads
\cite{no_p}
\begin{eqnarray}
{\cal Z} &=& 2^{N_{h}- N_{s}} \sum_{\eta}  {\cal Z}_\eta \nonumber \\ 
{\cal Z}_\eta &=& {\rm Tr} \exp[-\beta H_{K_h,\eta}],
\label{Zh}
\end{eqnarray}
where $N_{h} = \frac{N_{s}}{2}$ is the number of hexagonal plaquettes.
In terms of the original spins of Eq. (\ref{H})
\begin{eqnarray}
{\cal Z} = {\rm Tr} \sum_{n=0}^{\infty} \frac{(\beta
H_{K_{h}})^{2n}}{(2n)!},  
\label{even}
\end{eqnarray}
or, equivalently, in terms of the Majorana representation of
Eqs.(\ref{ha},\ref{he})
\begin{eqnarray}
{\cal Z}_\eta = {\rm Tr} \sum_{n=0}^{\infty}  \frac{(\beta
H_{K_h,\eta})^{2n}}{(2n)!}. 
\label{even1}
\end{eqnarray}

The reason why in Eq. (\ref{even}) we keep only keep the even powers of
$H_{K_{h}}$ is the following. By time reversal symmetry (due to the
trace over $\sigma^\mu_{i}$ and ($-\sigma^\mu_{i})$), at any given site
$i$ we must have an even power of $\sigma^\mu_{i}$. Similarly, in the
Majorana fermion representation [Eq. (\ref{even1})], the trace of
$c_{i}$ is zero. For any  odd power of $H_{K_{h}}$, there is in any 
term resulting from the expansion of $\exp[-\beta H_{K_{h}}]$ at least
one site for  which we have an odd power of $\sigma^\mu_{i}$ (or
$c_{i}$). All of these  terms vanish.

Let us first consider the Majorana representation and focus on ${\cal
Z}_\eta$. We will later on  rederive these results within the original
spin representation of Eq. (\ref{H}).  The local assignments
$\{{\eta}_{ij}\}$ effectively relate $J_{ij}$ in a general sector to
that in the sector $\{{\eta}_{ij} =1\}$ by the transformation
\begin{eqnarray}
J_{ij} {\eta}_{ij} \leftrightarrow  J_{ij}.
\end{eqnarray}
We claim that if a particular bond ($J_{ij}$) appears as an odd power in
a given term then it will give rise to a vanishing contribution  when it
is traced over. The proof of this  assertion is trivial:
\begin{eqnarray}
\sum_{{\eta}_{ij} = \pm 1} \eta_{ij}^{p} J_{ij}^{p} =0
\end{eqnarray}
for all odd $p$. 

The same conclusion follows within the original spin representation of
Eq. (\ref{H}) which as we show below leads to Eq. (\ref{even}). Let us
mark all the bonds $J_{ij}$ that would additionally appear to an odd
power in the  expansion of Eq. (\ref{even}). We claim that there are
several  possible topologies:

(i) three odd bonds (odd powers of $J_{ij}$) meet at a common vertex.

That is, we can have 
\begin{eqnarray}
J_{ij}^{p_{ij}} J_{ik}^{p_{ik}} J_{il}^{p_{il}}
\end{eqnarray}
with odd $p_{ia}$ ($a=j,k,l$) and with all of the bonds that touch $j,k$
and $l$ appearing to an even power in the expansion of Eq. (\ref{even}).

(ii) Closed or open contours of odd powered bonds appear.

In case (i), the spins at sites $j,k$ and $l$ appear to an odd power
(the power is just the sum of the powers of the bonds that have one of
these points at their end). In case (ii), if the contour is open then
the spins at the endpoints of the open contour must appear to an odd
power and therefore leads to a term that vanishes upon taking the
trace.  If the contour is closed it leads to none other than the anyon
charge within the contour $C$, 
\begin{eqnarray}
I_{C} = \prod_{h_\alpha \in C} I_{h_\alpha} 
\end{eqnarray}
with $I_{h_\alpha}$ the product of bonds along a  hexagonal loop. Using
the relation  $\sigma^{\mu} \sigma^{\nu} = i \epsilon_{\mu\nu\kappa}
\sigma^{\kappa}$ for $\mu \neq \nu$, we find that each spin $i$ that
lies on $C$ ($ i \in C$) leads to a contribution $\sigma^{\kappa}_{i}$.
For any odd power $p$, $[\sigma^{\kappa}_{i}]^{p} $ has a vanishing
trace. 

We can similarly, have both (i) and (ii). It is readily seen that all
odd powered bonds lead to  situations with either odd powers of the
spins at the endpoints and/or to closed contours which also lead to
vanishing contribution. 

Returning to the sum of Eqs.(\ref{Zh}, \ref{even1}), we now have 
\begin{eqnarray}
{\cal Z} = 2^{N_{h} - N_{s}+1} {\mathbf{E}} [ {\cal Z}_1] , 
\label{ze}
\end{eqnarray} 
where \cite{no_p} 
\begin{eqnarray}
{\cal Z}_1 = \exp[- \frac{\beta}{2} {\rm Tr} [M_1]] 
\ |\det(1+\exp(\beta N_1))|^{1/2}
\end{eqnarray}
is the partition function of the system in which all $\eta_{ij} = 1$
(the anyon free system).  In Eq. (\ref{ze}), the operator ${\mathbf{E}}$
projects out of ${\cal Z}_1$ only those terms that are even (hence the 
symbol ${\mathbf{E}}$) in all exchange constants $\{J_{ij}\}$.   The
matrices $N_1$ and $M_1$ depend on the constants $\{J_{ij}\}$ as
detailed in \cite{no_p}. The same conclusion follows from the expansion
Eq. (\ref{even}).

We note that for the particular set of exchange constants $J_{i, i+
\hat{e}_{\mu}} = J_{\mu}$, (nearest neighbor $J_{ij}$), Eq. (\ref{ze}) is
indeed equal to 
\begin{eqnarray}
{\cal Z} = 2^{N_{h}+1} {\mathbf{E}}^{z}[{\cal Z}_1] 
\end{eqnarray}
where ${\mathbf{E}}^{z}$ projects out of ${\cal Z}_1$ only the terms that
have all of the powers of $J_{i,i+\hat{e}_{z}}$ being even.  In order to
cover all of the topological sectors  $I_{h_\alpha} = \pm 1$, in each
hexagon $h_\alpha$ it suffices to allow ${\eta}_{ij} = \pm 1$ on all of
the vertical bonds (parallel to the $z$ direction), and ${\eta}_{ij} =1$
on all other bonds (parallel to the $x$ or $y$ directions)
\cite{nussinov_chen}. 

It is worth emphasizing that different topological sectors 
$\{I_{h_\alpha}\}$ lead to different ${\cal Z}_\eta$ (and thus to
different spectra as they indeed must). It is only after performing the
trace in Eq. (\ref{even1}) that the common even powered terms are pulled
out. These terms are the same in  all $\{{\eta}_{ij}\}$ assigments. 

\section{Vector exchange lattice model}

\subsection{Motivation} 

Consider the Lagrangian density of fermions coupled to a vector gauge
field $A_{a}$ with $a=0,1,2,3$. In a U(1) theory, the Lagrangian density
describing the minimal coupling of fermions to the gauge field is given
by
\begin{equation}
{\cal{L}}_{\min} = \bar{\psi}(i \gamma^{a} \partial_{a}- \gamma^{a} A_{a})
\psi, 
\label{minc}
\end{equation}
where $\gamma^{a}=(\gamma^{a})^\dagger$ are the Dirac matrices.  Within
the $U(1)$ theory, $A_{a = 0}$ is the scalar potential and $A_{a=1,2,3}$
are the spatial components of the  vector potential $\vec{A}$. The
minimal coupling term of Eq.(\ref{minc}) is augmented by a gauge-only
term ($\frac{1}{4}F_{ab} F^{ab}$ with $F_{ab} = \partial_{a} A_{b} -
\partial_{b} A_{a}$). In the electroweak theory ($SU(2) \times U(1)$),
the $A_{a}$ in Eq.(\ref{minc}) is replaced by $(A_{a} - V_{a})$ with the
weak parity breaking field $V_{a}$.

\begin{figure}[h]
\includegraphics[width=3.4in]{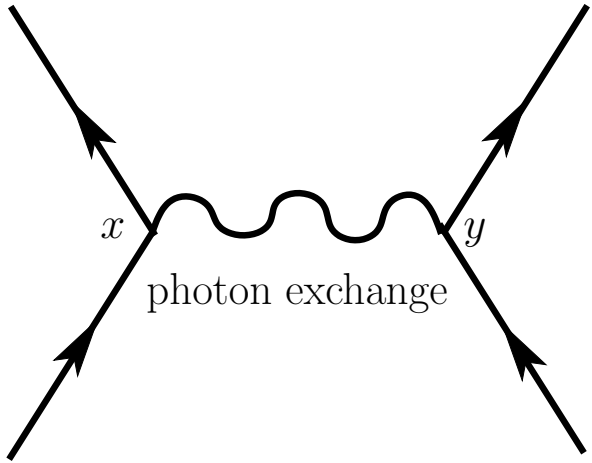}
\caption{Vector exchange between two fermions. To lowest (quadratic)
order in the gauge field, a bilinear between the Dirac matrices:
$\bar{\psi}(x) \gamma^{a} \psi(x) D_{a b}(x,y) \bar{\psi}(y) \gamma^{b} 
\psi(y)$ results. For a vector exchange ({\it photon}) propagator
$D_{ab}(x,y)$ set to be $J_{x,y}$ on a lattice, the resulting system is
precisely of the form of the vector exchange model of
Eq.(\ref{ourold}).}
\label{feyn_graph}
\end{figure}

Although the Lagrangian is quadratic in the fermion fields $\psi$, it is
definitely not a simple quadratic form that can be exactly integrated
out. This is due to the linear coupling to $A_a$. The theory contains
both {\it free} quadratic terms (e.g., those in $\psi$ alone) and terms
of the form $\bar{\psi} \gamma^a A_a \psi$. These terms give rise to
interactions such as the lowest-order exchange term shown in
Fig.(\ref{feyn_graph}). The coupling to the $A_a$ gauge field gives rise
to the usual Coulomb interaction between fermions. 

The lowest-order interaction terms are those formed by two vertices as
above. The slanted lines depict the fermions ($\psi$) while the
horizontal wavy line represents the {\it photon} propagator ($D_{ab}$)-
the propagator for the fields $A_{a}$. Integrating out the gauge field
gives rise to the usual Coulomb exchange (depicted in 
Fig.(\ref{feyn_graph})) 
\begin{equation}
\bar{\psi}(x) \gamma^{a} \psi(x) D_{ab}(x,y) \bar{\psi}(y) \gamma^{b} 
\psi(y)
\label{gg}
\end{equation}
with
\begin{equation}
D_{ab}(x,y) = \langle A_{a}(x)  A_{b}(y)  \rangle
\label{Coul}
\end{equation}
the Coulomb propagator. The same formalism albeit with more indices
applies to other vector gauges (e.g., the electroweak one).  In the
non-relativistic limit, the density-density interaction (the
$\bar{\psi}(x) \gamma^{0}  \psi(x) \bar{\psi}(y)  \gamma^{0} \psi(y)$)
piece becomes important. That is, the $D_{00}$  propagator becomes
dominant for non-relativistic particles.

The lattice gauge action for the fermions resulting   from integrating
out the vector gauges $A_{a}$ is not usually investigated in lattice
gauge theory calculations.  It is correct but this is not the standard
point of departure for lattice gauge calculations.  What is typically
done is to write terms of  the form
\begin{equation}
\bar{\psi}(x)  \gamma^{a} \psi(x+ \hat{e}_{a}) \ \exp[i
A_{x,x+\hat{e}_{a}}]
\end{equation}
with $A_{x,x+\hat{e}_{a}}$ the line-integrated lattice gauge (3+1)
vector potential  between nearest-neighbor sites.

What we do in the following affords another way of investigating
general minimally coupled actions.  When integrating out the $A_a$ fields,
we generate precisely interactions of the $\gamma \gamma$ type of
Eq.(\ref{gg}) with Eq.(\ref{Coul}). In what follows, we will investigate
a simple lattice rendition of such vector exchange system in which we
set an exchange coupling between $\gamma$ matrices to be of amplitude
$J_{ij} \equiv D(i,j)$  with $i$ and $j$ denoting lattice sites. 

\subsection{Exact solution of the vector exchange model}

\begin{figure}[h]
\includegraphics[width=3.4in]{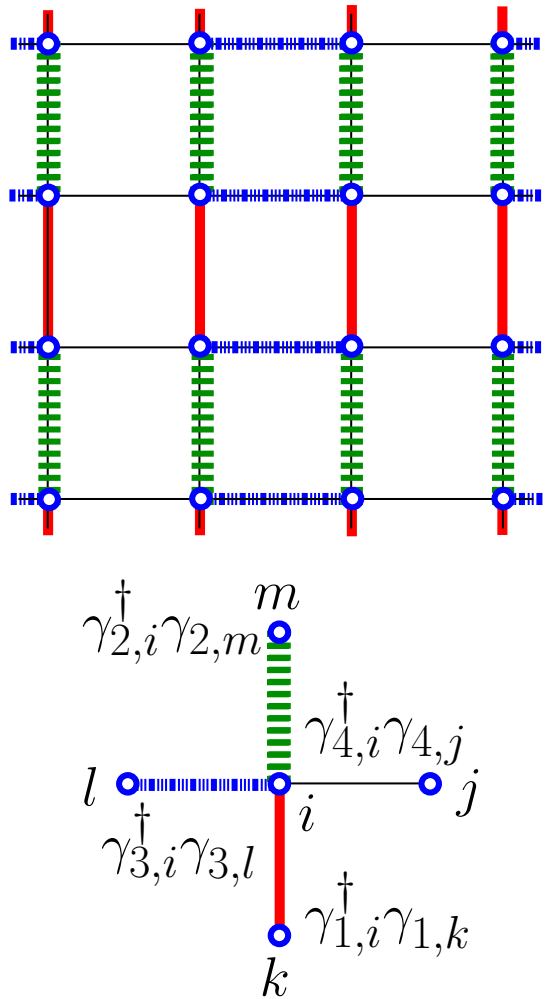}
\caption{The bonds in the system of  Eq.(\ref{ourold}). At each vertex
$i$ there are four different types of bonds corresponding to
$\gamma_{a,i} \gamma_{a,j}$ interactions with different $a=1,2,3,4$. 
(There are four different types of vertices.) The algebra of the bonds
is trivial: bonds that share a site anticommute, disjoint bonds commute,
and the square of any bond is one. Consequently the spectrum of the
model can be immediately determined.}
\label{figve}
\end{figure}

A simple square lattice model that captures the fermionic vector
exchange is given by 
\begin{eqnarray}
H = \sum_{\langle i j \rangle} J_{ij}  \gamma_{a,i} \gamma_{a,j}.
\label{ourold}
\end{eqnarray}
The geometry of the lattice is shown in Fig.(\ref{figve}). The $\gamma$
matrix index  $a$ for a given bond in Eq.(\ref{ourold}) is fixed by the
sites $i$ and $j$.  Here and throughout, we set the lattice constant 
to be one.

The $\gamma$ matrices satisfy the algebra
\begin{eqnarray}
\{ \gamma_{a,i}, \gamma_{b,i}\} = 2 \delta_{ab} \ , \ [\gamma_{a,i},
\gamma_{b,j}] = 0 , \ i \neq j.
\label{Cliff}
\end{eqnarray}
The Hilbert space on which $H$ acts on is, for a lattice of $N_s$ sites,
of dimension $4^{N_s}$. The algebra of the bonds $\gamma_{a,i}
\gamma_{a,j}$  is familiar: it has the same simple characteristics of
the bond algebra in Kitaev's honeycomb model. These algebraic relations
do not change on projection to a state of fixed topological charge
sector
\begin{eqnarray}
\hat{P}_{\Box}(\gamma_{a,i} \gamma_{a,j}) \hat{P}_{\Box} . 
\end{eqnarray}
We define the projector $P_{\Box}$  to a topological sector  by
\begin{eqnarray}
\hat{P}_{\Box} = \frac{\one + \tilde{I}_{\Box} I_{\Box}}{2} ,
\end{eqnarray}
with \begin{eqnarray}
I_{\Box} = \prod_{\langle ij \rangle \in \Box} \gamma_{a,i} \gamma_{a,j}
\label{plp}
\end{eqnarray}
As in Eq.(\ref{proj}), $\tilde{I}_{\Box}$ are $c$-numbers: 
$\tilde{I}_{\Box} = \pm 1$, and products of bonds around a plaquette 
$I_{\Box}$ commute with the Hamiltonian ($[I_{\Box}, H]= 0$), and
amongst themselves ($[I_{\Box}, I_{\Box'}]=0$). The operators of
Eqs.(\ref{plp}) constitute local (gauge) symmetries. The origin of the
commutation relations is that at each vertex we have bonds of different
$\gamma$ matrix flavors. Similar to the situation in Kitaev's honeycomb
model,  all bonds commute with the anyon charge operators of
Eq.(\ref{plp}), and  
\begin{eqnarray}
I_{\Box}^{2} =1.
\end{eqnarray}
The gauge symmetries $\{I_{\Box}\}$ allow decomposition of the total
Hilbert space into orthogonal subspaces of equal dimensionality. 
We divide the Hilbert space into equal sectors spanned by
$\{\tilde{I}_{\Box}\}$. There are $2^{N_s-1}$ such sectors as the eigenvalues of
$I_{\Box}$, for each of the $N_s$ plaquettes $\Box$, can attain one of
two values ($\pm 1$), and satisfy only one global constraint on a torus
\begin{eqnarray}
\prod_{\Box} I_{\Box} =  1.
\end{eqnarray}
As $\{\tilde{I}_{\Box}\}$ are good quantum numbers, we may diagonalize
the Hamiltonian in a Hilbert space of dimension $4^{N_s}/2^{N_s-1} =
2^{N_s+1}$. Similar to our solution of the Kitaev's honeycomb model, we
may then work with the representation of the bonds as the product of
two fermions. 

Within each anyon charge sector, the Hamiltonian is of the form of
Eq.(\ref{he}) but on different size spaces. We can now introduce $N_s$
spinless fermion variables $\{d_{i}\}$ on the Hilbert space of size
$2^{N_s}$ by setting the bonds to be 
\begin{eqnarray}
\bar{A}_{ij} = i (d_{i}+ d_{i}^{\dagger}) (d_{j}+ d_{j}^{\dagger}).
\label{fermimap}
\end{eqnarray}
We thus arrive at a Fermi bilinear that is trivially diagonalizable
\begin{eqnarray}
H =  i \sum_{\langle i j \rangle} \eta_{ij} J_{ij} (d_{i} +
d_{i}^{\dagger}) (d_{j} + d_{j}^{\dagger}).
\label{Hferm}
\end{eqnarray}
In Eq.(\ref{Hferm}), we maintain the directionality
that we employed throughout in constructing 
the bond algebra in the case of Kitaev's
honeycomb model: $j -i = \hat{e}_{x}$ or $\hat{e}_{y}$.
A trivial but important feature of Eq.(\ref{Hferm}) is that the
spectrum is symmetric about zero. This is so as there is a symmetry
$\bar{A}_{ij} \to - \bar{A}_{ij}$ in the representation chosen 
in Eq.(\ref{fermimap}).

The dimension of the Hilbert space is the same as that of the  product
of all plaquette charges $\tilde{I}_{\Box} = \prod_{\langle ij \rangle
\in {\Box}} \eta_{ij}$ (there are $(N_s-1)$ such Ising type operators
with eigevalues $\pm 1$ leading to $2^{N_s-1}$ sectors multiplied by the
size of the Hilbert space spanned by the $N_s$ fermions in a space of
size $2^{N_s}$ multiplied by a degeneracy factor of two. Fixed anyon
charges enable $2^{N_s}$ possible configurations  (redundant) of
$\eta_{ij}$ that give rise to the same  original Hamiltonian when
projected onto a sector of fixed $\{I_{\Box}\}$. These configurations of
$\eta_{ij}$ are related to each other by local Ising gauge
transformations on the square lattice. That is,  with arbitrary
$\tau_{i} = \pm 1$ at any lattice site $i$, the local gauge
transformation 
\begin{eqnarray}
\eta_{ij} \to \tau_{i} \eta_{ij} \tau_{j}
\end{eqnarray}
leaves $\tilde{I}_{\Box}$ invariant.

Written longhand, the $4^{N_{s}}$ dimensional Hilbert space spanned by
the $\gamma$ matrices decomposes as follows,
\begin{eqnarray}
4^{N_{s}} &= & [2^{N_{s}-1} \mbox{(number of sectors}~ \{
\tilde{I}_{\Box} \} ) \nonumber \\  
&\times& 2^{N_{s}} \mbox{(Hilbert space spanned by $N_{s}$ fermions)}
\nonumber \\  
&\times& 2 (\mbox{remaining degeneracy of each state})].
\end{eqnarray}
There is a degeneracy factor of $2= 4^{N_s}/(2^{N_s-1} \times
2^{N_s})$   that remains after invoking the representation of
Eq.(\ref{fermimap})  in the space of size $2^{N_s}$. This is similar to
the degeneracy factor of two in subsection \ref{alki}.
The Hamiltonian of Eq.(\ref{Hferm}) is nothing but a tight-binding
Hamiltonian agumented by pairing terms (an element of the so$(2N_s)$
algebra) on which we may apply a Bogoliubov transformation similar to
\cite{nussinov_chen}  which was defined on the square lattice. The
solution to Eq.(\ref{Hferm}) can be immediately written down. For 
$J_{ij}$ equal to $J_{x}$ or $J_{y}$ for sites $i$ and $j$ separated by
one lattice constant along the $x$ or $y$ directions respectively, i.e.
$J_{ij} = (J_{x} \delta_{|i_{x}-j_{x}|,1} +  J_{y} \delta_{|i_{y}-
j_{y}|,1}) $, in the  sector $\eta_{ij}=1$ (corresponding to
the sector $\tilde{I}_\Box= 1$), we have on 
Fourier transforming,
\begin{eqnarray}
H = i \Big[ \sum_{\vec{k}} q_{\vec{k}} d_{\vec{k}}^{\dagger} d_{-\vec{k}}^{\dagger}
+ q_{-\vec{k}} d_{\vec{k}} d_{-\vec{k}} \Big] \nonumber
\\ - \sum_{\vec{k}} p_{\vec{k}} d_{\vec{k}}^{\dagger} d_{\vec{k}} ,
\label{triv_k}
\end{eqnarray}
with $q_{\vec{k}} \equiv [J_{x} e^{ik_{x}} + J_{y} e^{ik_{y}}]$
and $p_{\vec{k}} \equiv 2 (J_{x} \sin k_{x} + J_{y} \sin k_{y})$.
A Bogoliubov transformation gives the quasi-particle spectrum
\begin{eqnarray}
E_{\vec{k}} = 0, \pm 2 p_{\vec{k}},
\end{eqnarray}
with the zero eigenvalue being doubly degenerate. 

\subsection{String correlation functions from symmetries}

The only correlators that can obtain a finite expectation value at zero
and finite temperatures must, by Elitzur's theorem \cite{Elitzur},  be
invariant under all local symmetries.  In \cite{nussinov_chen}, this
only allowed for string correlators of the form of Eq.(\ref{stringt}) to
have non-vanishing expectation values at finite and zero temperatures. 
These symmetry only conditions did not need to invoke the Majorana
fermion representation used in \cite{bas}. The considerations of
\cite{nussinov_chen} for Kitaev's model can be replicated mutatits
muntandis for the vector exchange model. The local  symmetries of
Eq.(\ref{plp}), allow only for string correlators (whether open or
closed) to attain a finite expectation value. This applies for both the
ground state configuration as well as the more physically pertinent case
of all finite temperatures, $T>0$. The zero temperature selection rule
was also noted by \cite{wu}.  Closed loops of the form of
Eq.(\ref{stringt}) that span the entire system correspond to additional
symmetries of the system. Relying on similar symmetry conditions as in
\cite{NO2}, it is seen that the only orders that can exist are of a
non-local nature. 

\subsection{Isomorphic spin models}

We now discuss two spin representations of our exactly-solvable vector
exchange model:

(i) a spin $S=1/2$ variant on a square lattice bilayer and 

(ii) a spin $S=3/2$ on a square lattice. This latter variant was very
recently also discussed by \cite{Yao,wu}. 
 
\subsubsection{A spin 1/2 model on a square lattice bilayer}
\label{ourmapsec}

A possible representation of Eq.(\ref{ourold}) is obtained at by
setting 
\begin{eqnarray}
\gamma_{\mu} = \left(
\begin{array}{cc}
0 & (-i \sigma^{\mu}) \\
(i \sigma^{\mu}) & 0 \\
\end{array}
\right), \ \gamma_{4} =  \left(
\begin{array}{cc}
\one_{2} & 0 \\
0 & -\one_{2} \\
\end{array}
\right),
\end{eqnarray}
with $\mu= 1,2,3$, $\sigma^{\mu}$ the Pauli matrices, and  $\one_{2}$
the 2-dimensional unit matrix.  The $\gamma$ matrices can arise from the
tensor product of two $S=1/2$ spins at each lattice site.
\begin{eqnarray}
\gamma_{1,i} &=& \sigma^{y}_{1,i} \sigma^{x}_{2,i}, \nonumber \\ 
\gamma_{2,i} &=& \sigma^{y}_{1,i} \sigma^{y}_{2,i}, \nonumber \\ 
\gamma_{3,i} &=& \sigma^{y}_{1,i} \sigma^{z}_{2,i}, \nonumber \\ 
\gamma_{4,i} &=& \sigma^{z}_{1,i}.
\label{grep1}
\end{eqnarray}

\begin{figure}[h]
\includegraphics[width=3.4in]{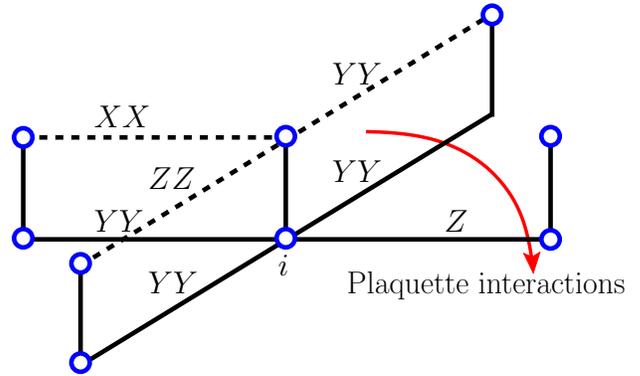}
\caption{A square lattice bilayer of a $S=1/2$ system that represents
Eq.(\ref{ourold}). Inserting Eq.(\ref{grep1}), we find the interactions
schematically depicted above. The two layers are indexed by the first of
the subscripts (1 or 2) in Eq.(\ref{grep1}). }
\label{planar}
\end{figure}

A possible lattice topology realization for this is  that of a square
lattice bilayer shown in Fig.(\ref{planar}). The subscript $\alpha$ in
$\sigma^{\mu}_{\alpha, i}$ is the bi-layer index.  Inserting
Eq.(\ref{grep1}) into Eq.(\ref{ourold}) leads to a spin $S=1/2$
Hamiltonian on a square lattice bilayer.

\subsubsection{A spin 3/2 model on a square lattice}

Another representation of the model of Eq.(\ref{ourold})  is in terms of
$S=3/2$ spins that reproduces the results of \cite{Yao,wu}  for
their  $J_{5}=0$. The {\it spin liquid} character of these systems is
inherited from Kitaev's honeycomb model.
\begin{eqnarray}
S^{x} = \left(
\begin{array}{cccc}
0 & \frac{\sqrt{3}}{2} & 0 & 0 \\
\frac{\sqrt{3}}{2} & 0 & 1 & 0 \\
0 & 1 & 0 & \frac{\sqrt{3}}{2} \\
0 & 0 & \frac{\sqrt{3}}{2} & 0 \\
\end{array}
\right),
\end{eqnarray}
\begin{eqnarray}
S^{y} = \left(
\begin{array}{cccc}
0 & -i \frac{\sqrt{3}}{2} & 0 & 0 \\
\frac{\sqrt{3}}{2} i & 0 & -i & 0 \\
0 & i & 0 & - \frac{\sqrt{3}}{2} i\\
0 & 0 & \frac{\sqrt{3}}{2} i & 0 \\
\end{array}
\right),
\end{eqnarray}
and 
\begin{eqnarray}
S^{z} = 
\left(
\begin{array}{cccc}
\frac{3}{2} & 0 & 0 & 0 \\
0 & \frac{1}{2} & 0 & 0 \\
0 & 0 & - \frac{1}{2} & 0 \\
0 & 0 & 0 & - \frac{3}{2}\\
\end{array}
\right).
\end{eqnarray}

In terms of two spins of size $S=1/2$,
\begin{eqnarray}
S^{x} &=& \frac{\sqrt{3}}{2} \sigma^{x}_{2} + \frac{1}{2}
(\sigma_{1}^{x} \sigma^{x}_{2} + \sigma_{1}^{y} \sigma_{2}^{y}),
\nonumber \\ 
S^{y} &=& \frac{\sqrt{3}}{2} \sigma^{y}_{2} + \frac{1}{2}
(\sigma_{1}^{y} \sigma_{2}^{x} - \sigma_{1}^{x} \sigma_{2}^{y}),
\nonumber \\ 
S^{z} &=& \sigma_{1}^{z} + \frac{1}{2} \sigma_{2}^{z},\nonumber \\
(S^{x})^{2} &=& \frac{\sqrt{3}}{2} \sigma_{1}^{x} - \frac{1}{2}
\sigma_{1}^{z} \sigma_{2}^{z}  + \frac{5}{4}, \nonumber \\
(S^{y})^{2} &=& - \frac{\sqrt{3}}{2} \sigma_{1}^{x}  - \frac{1}{2}
\sigma_{1}^{z} \sigma_{2}^{z}  + \frac{5}{4}, \nonumber \\ 
(S^{z})^{2} &=& \sigma_{1}^{z} \sigma_{2}^{z} + \frac{5}{4}, \nonumber
\\ 
\{S^{x}, S^{y}\} &=& \sqrt{3} \sigma_{1}^{y} , \nonumber
\\ 
\{S^{y}, S^{z}\} &=& \sqrt{3} \sigma_{1}^{z} \sigma_{2}^{y}, \nonumber
\\ 
\{S^{x}, S^{z}\} &=& \sqrt{3} \sigma_{1}^{z} \sigma_{2}^{x}.
\end{eqnarray}

We can represent the $\gamma$ matrices by 
\begin{eqnarray}
\gamma_{1} &=& \sigma_{1}^{z} \sigma_{2}^{y} = \frac{1}{\sqrt{3}}
\{S^{y}, S^{z}\}, \nonumber \\ 
\gamma_{2} &=& \sigma_{1}^{z} \sigma_{2}^{x} = \frac{1}{\sqrt{3}}
\{S^{x}, S^{z}\}, \nonumber \\ 
\gamma_{3} &=& \sigma_{1}^{y} = \frac{1}{\sqrt{3}} \{S^{x}, S^{y}\},
\nonumber \\ 
\gamma_{4} &=& \sigma_{1}^{x} = \frac{1}{\sqrt{3}} (S^{x})^{2} -
(S^{y})^{2}), \nonumber \\ 
\gamma_{5} &=& - \gamma_{1} \gamma_{2} \gamma_{3} \gamma_{4} =
\sigma_{1}^{z} \sigma_{2}^{z} =  (S^{z})^{2} - \frac{5}{4}.
\end{eqnarray}

\section{Clifford Algebraic Models}

The commonality of all these exactly-solvable models is the presence of
degrees of freedom that satisfy a Clifford algebra on-site
\begin{eqnarray}
\{\gamma_{a,i},\gamma_{b,i}\}=2\delta_{ab} ,
\end{eqnarray}
and a commutative algebra off-site
\begin{eqnarray}
[\gamma_{a,i},\gamma_{b,j}]=0 \ , \ i\neq j ,
\end{eqnarray}
with $a,b=1,\cdots,p$. The exactly-solvable Hamiltonians are then
written as linear combinations of quadratic products of these $\gamma$
matrices. Regardless of the dimension of the representation of the
$\gamma$ matrices, the Hamiltonian is always an element of so$(2N_s)$(in
the examples worked out in this paper), and thus a GMFH. 

From the viewpoint of lattice connectivity, notice a fundamental
difference between Kitaev's honeycomb and the $\gamma\gamma$
(vector-exchange) models. The coordination of the honeycomb lattice is
$z_h=3$, while the one for the $\gamma\gamma$ lattice is $z_\gamma=4$.
This is the reason why one needs $p=3$ anticommuting (Pauli) matrices in
the first case, while $p=4$ anticommuting ($\gamma$) matrices are needed
in the second model. The relation between $p$ and the dimension of the 
matrix representation of $\gamma$ is the following: When $p=2q$ or
$p=2q+1$, the matrix representation can be of dimension $2^q$. This is
the reason why Kitaev used Pauli matrices ($q=1$) in his honeycomb
model, while we had used Dirac matrices ($q=2$) in the vector-exchange
model. 

It is indeed obvious how to generalize these ideas to generate new
exactly-solvable models of the so$(2N)$ type in arbitrary dimensions and
for arbitrary lattice coordination. The idea consists in writing bond
operators which are quadratic products of Clifford operators which are
anticommuting on the same {\it lattice site}. The cardinal $p$ of that
set of operators will define the $z$ of the lattice (its connectivity). 
For instance, suppose we want to have a lattice with $z=5$. Then, a
Shastry-Sutherland-like connectivity lattice will do the job \cite{ssl}.
Now, write down a Hamiltonian which is a linear combination of bilinears
of  5 anticommuting $\gamma$ matrices that act upon a Hilbert space of
dimension $4^{N_s}$. This model will be exactly solvable. In this way,
we can construct a new model in a cubic $z=6$ lattice with $p=6$
anticommuting matrices, or as we show now a triangular $z=6$ lattice
model with $\gamma$ matrices of dimension $2^3 \times 2^3$. 

\begin{figure}[h]
\includegraphics[width=3.4in]{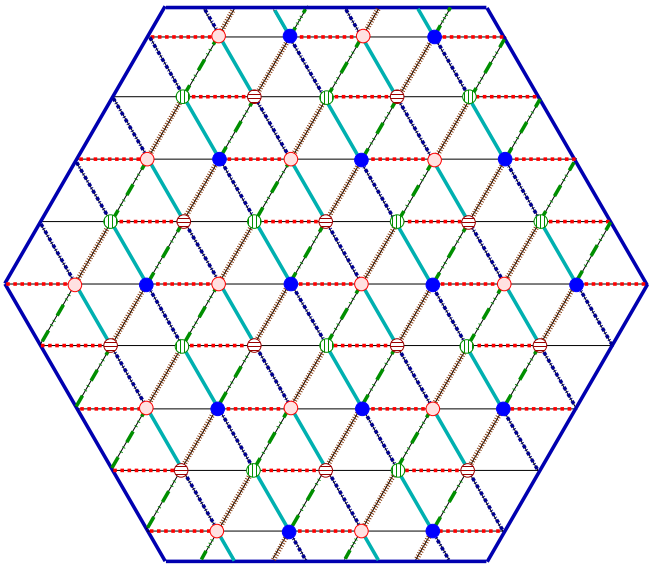}
\caption{A triangular lattice Clifford model that is exactly solvable.
In each vertex there is a degree of freedom of dimension $2^3$ (and a 
Hilbert space of dimension $2^{3N_s}$).  There are four different types
of vertices and six different types of bonds $\gamma_{a,i}
\gamma_{a,j}$.}
\label{figtri}
\end{figure}

Consider the $p=6$ $\gamma$ matrices
\begin{eqnarray}
\gamma_{1,i} &=& \sigma^{y}_{1,i} \sigma^{x}_{2,i}, \nonumber \\ 
\gamma_{2,i} &=& \sigma^{y}_{1,i} \sigma^{y}_{2,i}, \nonumber \\ 
\gamma_{3,i} &=& \sigma^{y}_{1,i} \sigma^{z}_{2,i}, \nonumber \\ 
\gamma_{4,i} &=& \sigma^{z}_{1,i} \sigma^{x}_{3,i}, \nonumber \\ 
\gamma_{5,i} &=& \sigma^{z}_{1,i} \sigma^{y}_{3,i}, \nonumber \\ 
\gamma_{6,i} &=& \sigma^{z}_{1,i} \sigma^{z}_{3,i},  
\label{grep2}
\end{eqnarray}
which form an on-site Clifford algebra. The model Hamiltonian
\begin{eqnarray}
H = \sum_{\langle i j \rangle} J_{ij}  \gamma_{a,i} \gamma_{a,j},
\label{ournew}
\end{eqnarray}
whose lattice geometry is shown in Fig.(\ref{figtri}), is exactly
solvable.
This model can also represent a tri-layer system with plaquette
interactions. 

One can indeed realize that there is nothing special about the Lie
algebra so$(2N)$. One can consider models whose bond algebra forms any
other semisimple Lie algebra, such as so$(2N+1)$ where there are
non-linear Bogoliubov transformations that diagonalize the problem. It
is important, though, that the number of gauge symmetries is enough to
allow for a simple oscillator realization of the bonds. Otherwise, there
is always the possibility to use the Jacobi algorithm \cite{solvability}
to numerically diagonalize the problem.  

\section{Conclusions}

The thesis of the current work is that even though the solution to many
problems is hard, by disregarding the explicit microscopic degrees of
freedom and focusing solely on the algebraic relations that the bond
variables satisfy in a Hilbert space of a fixed dimensionality,  we may
map one initially seemingly hard  problem onto another problem whose
solution is easier. This mapping does not rely on explicit real space
forms for the transformations (although these can be written down in
some cases). Nor does it rely on  enlarging the Hilbert space and then
making a projection onto a physical sector as in \cite{Kitaev2006}.  It
is important to emphasize, though, that the dimension of the  {\em
representation} of the algebra is of crucial importance - not only the
algebra and set of constraints itself. 

The explicit real space mappings -- no matter how complicated their
forms are -- are irrelevant.  The spectra and all non-vanishing 
correlators may be determined from the algebra alone. [See \cite{NO2}
for a derivation of all correlators in Kitaev's Toric code model by this
method.]   In the current work,  we illustrated how the energy spectra
may be determined.  The partition function and the density of states
associated with the spectrum are related by a Laplace transform.
It may be easily seen also from the partition functions themselves
that if two systems display the same bond algebra on a space of the same
representation then their spectra
are identical. 

In the case of generic GMFHs  the Jacobi method always enables a
solution of its spectrum  with polynomial complexity \cite{solvability}.
There are situations, like the model examples presented in this paper,
where we can  determine the spectrum of certain sectors (Hilbert
subspaces) in closed form, i.e., by quadrature. This will happen
whenever the effective matrices that need to be diagonalized have
dimension smaller or equal to  4 $\times 4$. The decomposition of the
Hilbert space into these individual decoupled subspaces is rooted in the
existence of {\em local (gauge) symmetries}. Within each subspace, there
is an  {\em oscillator realization} (e.g., in terms of Majorana
fermions) of the bond algebra.  It is important to emphasize that exact
solvability does not imply that we can compute the density of states,
and thus the partition function, with polynomial complexity. Kitaev's
honeycomb model is an example of a system whose energy eigenvalues can
be determined with polynomial complexity but whose total partition
function cannot since its density  of states is not determined with the
same complexity. By contrast, Kitaev's Toric code model constitutes an
example of a system where not only the spectrum but also its partition
function is exactly solvable \cite{NO1,NO2}. 

{\em Note added in proof. } This work (and our vector-exchange model) 
was conceived in 2007. A physical model whose exact solution was enabled
by our bond algebra mapping is detailed in \cite{orbit}. During the time
in which the current work was summarized, three works appeared
\cite{Yao,wu,ryu} that introduce and solve variants  and exact forms of
the vector-exchange model that we introduced here. In particular,
\cite{wu} presented a $S=3/2$ spin system and the vector-exchange model
on the Shastry-Sutherland \cite{ssl} decorated square lattice. 


\begin{thebibliography}{28}
\bibitem{longT} 
It is important to note that all of the steps invoked in the 
high-temperature expansion discussed below rely only on the operator
algebra and the dimension of its representation. First, note that
$\exp[\beta (A_{s} + B_{p})] = \exp(\beta A_{s}) \exp(\beta B_{p})$.
Then, the partition function is 
\begin{eqnarray}
{\cal Z} = (\cosh \beta)^{2N_s} {\rm Tr}_{\{\sigma\}}  \sum_{G}(1+ (\tanh
\beta )^{|S|} \prod_{s \in G} A_{s}) \nonumber \\ 
\times (1+ (\tanh \beta)^{|P|} \prod_{p \in G} B_{p}),
\label{zg}
\end{eqnarray} 
with $G$ all graphs on the lattice (subset of lattice sites). The graph
$G$ spans $|S|$ sites and $|P|$ plaquettes. As $A_{s} = \prod_{i}
\sigma^{x}_{is}$ and $B_{p} = \prod_{ij \in p} \sigma^{z}_{ij}$, and as
the trace of any Pauli matrix $\sigma^\mu_{ij}$ raised to an odd power
is zero, the only terms surviving in the sum of Eq.(\ref{zg}) are those
with graphs $G$ such that each lattice link appears in an even number of
star operators (i.e., zero or two) and an even number of plaquettes
(zero or two). The trace over these non-vanishing terms is $2^{2N_s}$,
as there are $2^{2N_s}$ possible bond configurations (and a total of
$2N_s$ bonds) and $\sigma^\mu_{ij}$ raised to an even power is the
identity operator.   The only graphs $G$ on a torus that  satisfy this
condition are those in which (i) $G$ is the empty set,  (ii) $G$
contains all stars in the lattice, (iii) $G$ contains all plaquettes in
the lattice, or (iv) $G$ contains all stars and plaquettes in the
lattice. These give rise to finite terms as the products over the entire
torus, $\prod_{s} A_{s} =1$ and $\prod_{p} B_{p} =1$, by virtue of the
topology (each bond on a square lattice on a torus  appears in an even
number of plaquettes or stars). Summing up these four  graphs we have,
Eq.(\ref{zend}).
\bibitem{crossover_trivial}
As a curiosity, it is worth noting that our bond algebraic mapping gives
rise to a crossover for finite size systems. As we showed by the
application of bond algebras, Kitaev's  Toric code model is isomorphic
to two decoupled Ising chains. The correlation length within an Ising
chain of exchange constant $J=1$ as in Eq.(\ref{h1d}) is given by
\begin{eqnarray}
\xi = \frac{1}{\ln(\coth \beta)}.
\end{eqnarray}
When $\xi \gg N_s$, the system may appear to be ordered while for $\xi
\ll N_s$,  the system is disordered across the chain. When $\beta \gg
1$, the function
\begin{eqnarray}
\frac{1}{\ln(\coth \beta) } \to \frac{e^{2 \beta}}{2}.
\end{eqnarray}
Consequently, the crossover temperature at which $\xi(T) \simeq N_s$ is
given by
\begin{eqnarray}
k_{B} T_{\sf cross} = \frac{2}{\ln(2N_s)}.
\label{tcross}
\end{eqnarray}
For $T<T_{\sf cross}$ the system may support instantaneous local order
across a finite size system. As $N_s \to \infty$, the crossover
temperature tends to zero. The existence of a crossover temperature with
a $1/ \ln(2N_s)$ scaling that we derived here (Eq.(\ref{tcross})) from our
old bond algebra mapping to a one-dimensional system \cite{NO1, NO2}
coincides with that seen in the entanglement entropy \cite{cc}. Related
scalings for mutual information were derived in \cite{pachos}.
\end{thebibliography}

\end{document}